\documentclass[10pt,aps,pre,amsmath,amsfonts,amssymb,floatfix,superscriptaddress,notitlepage,nofootinbib]{revtex4-1}
\usepackage{mathrsfs} \usepackage{graphicx,color} 
\usepackage{multirow}
\usepackage{bm}
\usepackage[normalem]{ulem}

   \let\d=\delta
   
\let\l=\lambda \let\m=\mu   
 \let\t=\tau  
   
\let\D=\Delta

\def\PP{{\cal P}} 
 \def\HH{{\cal H}}
 
  \def\OO{{\cal O}}

\def\de{\mathrm d}
\def\dt{{\mathrm d}t}

\def\to{\rightarrow}

\newcommand{\beq}{\begin{equation}} \newcommand{\eeq}{\end{equation}}

\begin{document}
\title{
Statistical physics of learning in high-dimensional chaotic systems
% of learning from chaos
}

\author{Samantha J. Fournier}
\affiliation{
Universit\'e Paris-Saclay, CNRS, CEA, Institut de physique th\'eorique, F-91191 Gif-sur-Yvette, France
}

\author{Pierfrancesco Urbani}
\affiliation{
Universit\'e Paris-Saclay, CNRS, CEA, Institut de physique th\'eorique, F-91191 Gif-sur-Yvette, France
}

\begin{abstract}
In many complex systems, elementary units live in a chaotic environment and need to adapt their strategies to perform a task, by extracting information from the environment and controlling the feedback loop on it. One of the main example of systems of this kind is provided by recurrent neural networks. In this case, recurrent connections between neurons drive chaotic behavior and when learning takes place, the response of the system to a perturbation should take into account also its feedback on the dynamics of the network itself. In this work, we consider an abstract model of a high-dimensional chaotic system as a paradigmatic model and study its dynamics. We study the model under two particular settings: Hebbian driving and FORCE training. In the first case, we show that Hebbian driving can be used to tune the level of chaos in the dynamics and this reproduces some results recently obtained in the study of more biologically realistic models of recurrent neural networks. In the latter case, we show that the dynamical system can be trained to reproduce simple periodic functions. To do this, we consider the FORCE algorithm –originally developed to train recurrent neural networks– and adapt it to our high-dimensional chaotic system. We show that this algorithm drives the dynamics close to an asymptotic attractor the larger the training time. All our results are valid in the thermodynamic limit thanks to an exact analysis of the dynamics through dynamical mean field theory.
\end{abstract}

\maketitle

\section{Introduction}
Biological neural networks can be described at a first approximation as elementary units, the neurons, which interact through synaptic connections. 
Neurons are non-linear response units in the sense that only if their incoming input current is larger than a threshold value, they spike an action potential which may trigger a spike train in other neurons \cite{kandel2000principles, dayan2005theoretical}. 
Such intermittent non-linear dynamics is at the fundamental basis of all high level brain activities and the way in which this micro-dynamics becomes the macro-response that triggers movements and actions in organisms is still not understood. 
However, it is believed that biological neural networks are not randomly connected. The synaptic connections between neurons are plastic and can be tuned (learned) to perform cognitive tasks. But the mechanism allowing such control and adaptation is still largely unknown \cite{abbott2000synaptic}.

This is at odds with artificial neural networks (ANNs) at the basis of the deep learning revolution. ANNs are high-dimensional networks typically trained to solve an optimization problem, be it to classify images \cite{krizhevsky2017imagenet}, denoise them \cite{elad2023image}, or generate synthetic images \cite{goodfellow2020generative, song2020score}. Generally, the feedforward structure of the architecture is very helpful since it allows the implementation of gradient based optimization algorithms, such as stochastic gradient descent through backpropagation. 

Conversely in loopy networks, such training strategies are much more hard to implement. When the output of the neurons can be fed back into the neurons themselves, gradient based algorithms may become unstable because feedback signals may amplify or diminish, leading to diverging or vanishing gradients and non-converging dynamics. Since recurrent neural networks (RNNs) are closer (to some extent) to biological neural networks, the training problem in this case has become central also as a benchmark to propose biologically inspired learning strategies, which may be tested at the level of neurons' interactions and biology.

Models of RNNs have been studied for a long time \cite{sompolinsky1988chaos}. In the simplest of settings, synaptic connections are random and no training is performed. In this case, one can observe that depending on the strength of the interactions between neurons, the dynamics can be either quiescent or chaotic. It has been shown through numerical simulations in the latter case that such RNNs can be successfully trained to perform a simple task \cite{sussillo2009generating}. This is done by considering a special subset of the network as a readout device whose output is fed back into the network to allow its control. Therefore, the training task aims at using the output device to suppress chaos and generate the desired response. 

It is fair to say that this framework applies not only to recurrent neural networks. Biological systems as well as other complex systems (the financial market for example) typically live in chaotic environments and adaptation can be seen as a way to extract information from the (high-dimensional) chaos, and to adapt and control the feedback loop on the environment itself. 
Therefore, how to control and learn in chaotic environments is an ubiquitous problem.
The purpose of this manuscript is to start the investigation of such problems in a simplified high-dimensional setting. 

Instead of looking at specific models of RNNs or other complex chaotic systems, we consider an abstract high-dimensional chaotic system. There are several reasons to perform this abstraction step: on the one hand, we will show that the phenomenology found in specific realistic models can be found also in abstract ones, showing some degree of universality. On the other hand, the abstract models we present here have the advantage to be simpler to study from the statistical physics point of view. In particular, the dynamical mean field theory (DMFT) that we present will provide a set of equations which describe the dynamics of the models in the thermodynamic limit. These equations can be integrated numerically more efficiently than in other systems.
Since our primary goal is the application of these abstract models to RNNs, we analyze them in two steps. First, we show that the class of models that we consider share the same phenomenology as standard RNNs when untrained. In particular, we show that they can have a quiescent-to-chaotic transition as a function of the interaction strength between the degrees of freedom \cite{sompolinsky1988chaos}, and that the level of chaos in the chaotic phase can be tuned by Hebbian driving, analogously to what has been found in RNN models \cite{clark2023theory}. This implies that the models we consider are perfectly equivalent from the collective dynamics point of view to RNNs. Second, 
Sussillo and Abbott  \cite{sussillo2009generating} have shown through numerical simulations that specific models of RNNs in their chaotic phase can be trained to perform a simple task. They developed an algorithm called FORCE to do this. We adapt their algorithm to our dynamical system and investigate the performances of this algorithm in the thermodynamic limit. We achieve this using DMFT. The main advantage of using our abstract models rather than standard RNNs (where the same analysis could be developed in principle) is that in order to study the learning dynamics, one needs to get access to long transient timescales, which is hard to do in standard RNNs where the DMFT analysis is much more complicated than in our case. The algebraic structure of our models is better suited for this task and therefore we manage to study learning in this case.

The plan of the paper is the following.
In Sec.\ref{dyn_sys_sec}, we will describe a simple set of high-dimensional chaotic dynamical systems which we will use as simplified abstract models. In Sec.\ref{Hebbian_training_sec}, we will discuss what happens when these dynamical systems are subjected to Hebbian driving, namely when the dynamics of the system itself shapes the synaptic interactions (with a simple form of the Hebb rule). In this case, we show that the abstract dynamical systems that we consider displays the same phenomenology that has been found in the context of a standard, more biologically inspired RNN under the same type of training. In particular, Clark and Abbott \cite{clark2023theory} have recently shown that Hebbian driving can shape chaos and suppress it, up to the point that the plastic synaptic couplings become so strong that chaos is completely frozen. We will review the phenomenology observed in \cite{clark2023theory} and develop a theory for it in the context of our simplified setting.
In Sec.\ref{FORCE_training_sec}, we will instead consider a proper learning strategy. We will follow Ref.\cite{sussillo2009generating} and add to the dynamical system a readout unit which has to be trained such that its output matches a desired one. In order to perform this task, we will consider the FORCE algorithm \cite{sussillo2009generating} and adapt it to our dynamical systems. We DMFT analysis to show that the algorithm is effective in training the system also in the infinite size limit, and we track the behavior of the dynamical system during learning as a function of time. We will show that the learning dynamics bring the system closer to a dynamical attractor the longer the training time.
Finally in Sec.\ref{perspectives_sec}, we will discuss some perspectives on how to extend our framework.

\section{A simple high-dimensional chaotic system}\label{dyn_sys_sec}
The simplest model of a recurrent neural network (RNN) is defined by a set of $N$ neurons identified by an index $i=1,\ldots, N$. The state of each neuron is described by two variables, its membrane potential $x_i$ and its firing rate $r_i$. The firing rate is in general a non-linear function of the membrane potential, typically $r_i=\tanh (x_i)$. The dynamics of the network is described by a set of ordinary differential equations
\beq
\dot x_i(t) = -x_i(t) +\frac{g}{\sqrt N}\sum_{j(\neq i)} J_i^j r_j(t) + H_i(t),
\label{oldRNN}
\eeq
where the dot denotes the derivative with respect to time.
Here, the matrix $J_i^j$ describes the interactions between different neurons. Most importantly, this matrix is not supposed to be symmetric and therefore we will assume that $J^j_i\neq J_j^i$. 
Finally, $H_i(t)$ models some input current in neuron $i$. 
The model in Eq.~\eqref{oldRNN} has been studied extensively in the past, especially when the synaptic coupling matrix $J$ is thrown at random and fixed. In the simplest setting, one can assume that $J_i^j$ are just independent Gaussian random variables with zero mean and unit variance.
The control parameter $g$ describes the strength of the random interactions between neurons.
In the absence of the external current $H_i$ and for $g=0$, the dynamics of the network is described by a single stable attractor where $x_i=0$ for all $i=1,\ldots, N$, meaning that all neurons are at rest. This attractor becomes unstable under linear perturbations as soon as $g>g_c$, where $g_c=1$. In this case, the dynamics of the network is chaotic and the transition to chaos has been studied extensively in the past, see the pioneering work by Sompolinsky, Crisanti and Sommers \cite{sompolinsky1988chaos} who developed the dynamical mean field theory for Eq.~\eqref{oldRNN}. In this chaotic phase, it has been shown in \cite{sussillo2009generating, sussillo2009learning}  through numerical simulations on a finite system that the neural network can be efficiently trained. Therefore in the following, we will mainly focus on the properties of the chaotic phase.

The purpose of this work is to investigate up to which point Eq.~\eqref{oldRNN} can be simplified, while retaining its main physical properties.
For $g>g_c$, Eq.~\eqref{oldRNN} represents a chaotic high-dimensional non-linear dynamical system.
Therefore, we consider a different model still described by a set of $N$ real dynamical variables $\underline x=\{x_i\}_{i=1,\ldots, N}\in \mathbb{R}^N$, but we avoid the introduction of the firing rates $r_i$ which complicate the DMFT analysis, see \cite{sompolinsky1988chaos}.
In order to introduce the non-linearity in the equation we assume that
\beq
\dot x_i(t) = -\mu(t)x_i + \frac{\hat g}{N}\sum_{j,k}J_i^{jk}x_j x_k +H_i(t).
\label{dyn_sys}
\eeq
The matrices $J_i$ are chosen to be GOE random matrices which means that
\beq
J^{jk}_i=J^{kj}_i
\eeq
and 
\beq
\overline{J_i^{jk}} = 0  \ \ \ \ \ \ \overline{\left(J_i^{j \neq k}\right)^2} = 1 \ \ \ \ \ \ \ \ \overline{\left(J_i^{jj}\right)^2} = 2\:.
\eeq
We also assume that the matrices $J_i$ and $J_{j(\neq i)}$ are independent and identically distributed. We emphasize that Eq.~\eqref{dyn_sys} has to be regarded as a non-linear, high-dimensional random dynamical system and the purpose of this paper is to investigate how much it resembles more standard models of RNNs. A similar dynamical system has been used to study driven glasses in \cite{berthier2000two}, and the main difference with our current approach is that in \cite{berthier2000two} one adds to the lhs of Eq.~\eqref{dyn_sys} a conservative random force term which we completely avoid.
Here, we would like to consider the model described by Eq.~\eqref{dyn_sys} as a simplified model of a RNN. Clearly, this model is not biologically plausible in the sense that the microscopic form of the dynamics is rather far from standard models such as Eq.~\eqref{oldRNN}, which try to model microscopic interactions between neurons. However, we will argue that the model has the same phenomenology as the more standard model of RNNs described by Eq.~\eqref{oldRNN}. The main reason to choose a dynamical system of the form of Eq.~\eqref{dyn_sys} is that it is simpler to study from the theoretical point of view. {In particular, when we will come to study learning dynamics, we will need to develop the DMFT analysis at large timescales and this is very difficult for standard models of RNN such as Eq.~\eqref{oldRNN}.}

We will study the behavior of the dynamical system described by Eq.~\eqref{dyn_sys} under different settings.
First in Sec.~\ref{Hebbian_training_sec}, we follow the recent work by Clark and Abbott \cite{clark2023theory} and introduce a Hebbian driving term in the dynamical system. We show that depending on the strength of the Hebbian couplings, one can either reduce the chaotic activity or freeze it completely to lead the network to a random fixed point attractor.
Second in Sec.~\ref{FORCE_training_sec}, we will discuss how Eq.~\eqref{dyn_sys} can be trained to reproduce a simple periodic function using the FORCE algorithm developed by Sussillo and Abbott in \cite{sussillo2009generating} and originally described to train the system in Eq.~\eqref{oldRNN}.
We will also consider the discrete time algorithm defined by the Euler discretization of Eq.~\eqref{dyn_sys}, defined as
\beq
x_i(t+\de t) = x_i(t)+\de t\left[ -\mu(t)x_i + \frac{\hat g}{N}\sum_{j,k}J_i^{jk}x_j x_k +H_i(t)\right].
\label{dyn_discrete}
\eeq
At variance with the continuous time dynamics, such dynamical system depends also on the learning rate $\de t$.
Both dynamical systems in Eq.~\eqref{dyn_sys} and \eqref{dyn_discrete} depend also on a confining potential term proportional to $\mu(t)$ which is enforced in order to avoid that the dynamics diverges to infinity.

In the following, we will develop a DMFT analysis which allows us to understand how the dynamical system behaves in the infinite size limit $N\to\infty$.

\subsection{The statistical properties of the chaotic term}
A crucial step to understand the behavior of Eq.~\eqref{dyn_sys} is to analyze the chaotic term defined by the random matrices $J_i$.
It is useful to study the statistical properties of this term
\beq
\xi_i(t) = \frac{\hat g}{N}\sum_{j,k}J_i^{jk}x_j(t) x_k(t)\:.
\label{micro_xi}
\eeq
It is clear that the average over the random matrix realization gives
\beq
\overline{\xi_i}=0\:.
\eeq
However $\xi$ has an interesting dynamical two point correlation function
\beq
\overline{\xi_i(t) \xi_j(t')} = 2 \hat g^2\delta_{ij} C^2(t,t'),
\eeq
where the correlation function $C(t,t')$ is defined as
\beq
C(t,t') = \frac{1}{N} \sum_{i=1}^N x_i(t)x_i(t')\:.
\eeq
Higher order correlation functions factorize and can be computed through Wick contractions
due to the Gaussian nature of the matrices $J_i$.

Finally, we note that the form of the chaotic term is a particular case of a more general form. Indeed one can generalize
\beq
\xi_i(t) = \sum_{q=1}^{\infty} \frac{c_q}{N^{q/2}} J_i^{j_1\ldots j_q} x_{j_1}(t)\ldots x_{j_q}(t)\:.
\label{generic_micro_xi}
\eeq
By tuning carefully the coefficients $c_q$, one can get 
\beq
\overline{\xi_i(t) \xi_j(t')} = \Xi(C(t,t')),
\label{generic_Xi}
\eeq
where $\Xi(z)$ is an arbitrary positive function for $z>0$.
In particular one can show that $c_q^2$ enters in the coefficient of the $q$-th term of the Taylor expansion of $\Xi(z)$.
Note that both Eq.~\eqref{micro_xi} and \eqref{generic_micro_xi} describe a multibody interaction potential term. This is certainly not so natural from the biological perspective. However, in this particular work we use a multibody interaction because it is trivial to see that if $\Xi$ is a linear function, the dynamical system becomes linear itself and therefore it is fully integrable if $\mu(t)$ does not depend on $\underline x$.

\subsection{The confining potential term}
Since the degrees of freedom in both Eq.~\eqref{dyn_sys} and \eqref{dyn_discrete} are continuous and real, one needs to enforce a confining mechanism to avoid that the system explores an infinite phase space.
In the following, we choose two options.
\begin{itemize}
\item A standard way to impose a compact phase space is to bound the norm of the vector $\underline x$. Without losing generality, we enforce 
\beq
\sum_{i=0}^N x_i(t)^2 = N
\label{spherical_con}
\eeq
and we dub the corresponding model as a \emph{spherical model}.
This implies that coupling $\mu(t)$ is self consistently determined to assure that at each infinitesimal time step the dynamical system never leaves the constraint in Eq.~\eqref{spherical_con}. We anticipate that in this case, the DMFT equations track the dynamics only in the continuous time limit, while the discrete time dynamics has a natural correction of order $\de t^2$ which is not properly taken into account by the Euler discretization of the DMFT equations \cite{sarao2021analytical,mignacco2022effective}. We also note that this form of the constraint is confining whatever the nature of the chaotic noise $\xi$ and the corresponding form of its correlation functions $\Xi$.
\item A different way to impose a confining potential is to consider a term that  penalizes wild fluctuations of the norm of $\underline x$. 
A simple way to do that is to consider \cite{sarao2021analytical}
\beq
\mu(t) = f\left[\frac{1}{N}\sum_{i=1}^N x_i(t)^2\right],
\label{mu_generic}
\eeq
where the function $f(z)$ is positive and diverging function for $z\to \infty$.
We dub the corresponding model a \emph{confined model}.
In this case, the DMFT dynamics can be tracked also in the discrete time step case \cite{mignacco2022effective}. However, the confining capability of the form in Eq.~\eqref{mu_generic} depends strictly on the nature of the chaotic noise. 
In particular if we assume that both $\Xi(z)$ and $f(z)$ admit a polynomial expansion of finite degree, which degree we indicate respectively as $d_\Xi$ and $d_f$, then the resulting dynamics is confined if
\beq
d_\Xi < d_f\:.
\label{constraint_noise}
\eeq
\end{itemize}

\section{Transition to chaos and Hebbian driving}
\label{Hebbian_training_sec}
We would now like to investigate whether the prototypical model of Eq.~\eqref{dyn_sys} is a qualitatively good model for RNNs. Specifically, we will focus on two aspects: first, we will show that the class of models in Eq.~\eqref{dyn_sys} can have a phase transition from a quiescent attractor phase to a chaotic activity phase, as the model in Eq.~\eqref{oldRNN}. Second, we will follow a recent work by Clark and Abbott \cite{clark2023theory} who showed that the level of chaos in a model of RNN described by Eq.~\eqref{oldRNN} can be tuned by Hebbian driving of synapses. We will show that we can recover the same phenomenology as in \cite{clark2023theory}  and we will analyze the corresponding dynamics in the thermodynamic limit through DMFT.

\subsection{Transition to chaotic dynamics}

\begin{figure}
\centering
\includegraphics[width=\columnwidth
]{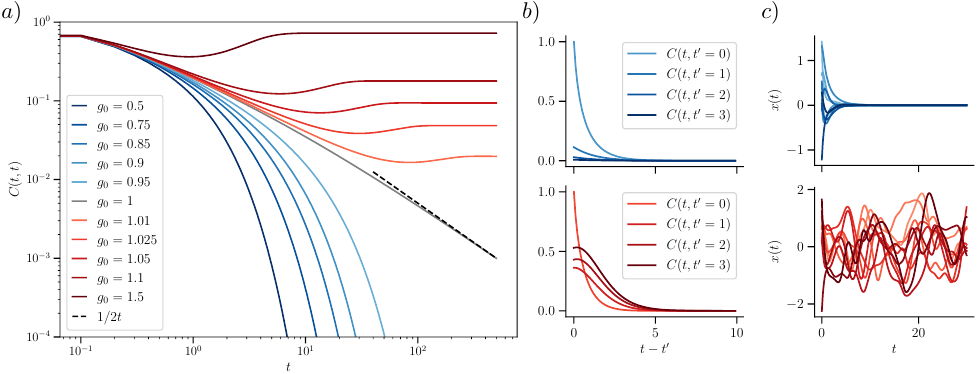}
\caption{Transition to chaos for the confined model defined by Eqs.~\eqref{mu_1} and \eqref{xi_1} with $g=1$ and $\mu(t)=1+C(t,t)$. $a)$ Behavior of $C(t,t)$. The system is randomly initialized such that $C(0,0)=1.$ For $g_0<g_0^c=1$, the dynamics is attracted by the fixed point $\underline x=0$. At the critical point $g_0=1$, the dynamics relaxes on the attractor with a power law decay. For $g_0>g_0^c$, the dynamics stays chaotic and stabilizes on a region of phase space characterized by a limiting value of the norm $|\underline x(t)|^2=\lim_{t\to \infty}C(t,t)$. $b)$ Behavior of $C(t,t^\prime)$ for different values of $t^\prime$ as a function of $t-t^\prime$, for $g_0=0.5$ (top) and $g_0=1.5$ (bottom). The chaotic regime (top) is characterized by fast decorrelation dynamics. $c)$ Some traces of $\underline{x}$ obtained through numerical simulations for $g_0=0.5$ (top) and $g_0=1.5$ (bottom). When $g_0=0.5$, all the $x_i$ go to $0$; while they display chaotic dynamics for $g_0=1.5$.}
\label{transition_to_chaos}
\end{figure}

In Sect. \ref{section: hebbian driving}, we will consider the spherical model with $\Xi(z) = 2\hat g^2 z^2$.
However in this case, given that the dynamics is constrained to be on the sphere and that there is no confining term pushing the system to a stable quiescent fixed point as in Eq.~\eqref{oldRNN}, one never encounters an attractor: the dynamics is always driven by the chaotic term whatever the strength of $\hat g$, as far as $\hat g>0$.
Therefore –at variance with the more standard model in Eq.~\eqref{oldRNN}– the present model lacks a phase in which the dynamical system goes at rest to a stable attractor.
In order to study this case, we consider a slightly different model, namely a confined model with
\beq
\mu(t) = 1 + C(t,t)\:.
\label{mu_1}
\eeq
Furthermore, we choose the following form for the correlation of the chaotic noise term
\beq
\Xi(z) = g_0^2z+ \frac{3 g^2}{2}z^2
\label{xi_1}
\eeq
and this corresponds to have a noise term of the form
\beq
\xi_i(t) = \frac{g_0}{\sqrt N}\sum_{j=1}^N J_i^j x_j(t) + \frac{\hat g}{N}\sum_{jk}^N J_i^{jk}x_j(t) x_k(t),
\eeq
with $2\hat g^2=3g^2/2$.
We are interested in considering what happens to the dynamical system as a function of $g_0$ at fixed $g$. We assume that the dynamics starts from an initial condition that is drawn from the flat measure over the sphere $C(t,t)=1$.
For $g_0=0$, the dynamical system has a fixed point at $\underline x=0$ and a random initialization of the dynamics leads to this fixed point, see Fig.\ref{transition_to_chaos}. 
As for the neural network in Eq.~\eqref{oldRNN}, one can have a chaotic transition as a function of $g_0$. This happens when the fixed point at the origin looses linear stability. 
Indeed, by linearizing the dynamical system around $\underline x=0$, one sees that the dynamics is described by $\delta\dot{\underline x}(t) = \HH \delta \underline x(t)$, with the matrix $\HH_{ij} = -\delta_{ij}+g_0J^j_i/\sqrt{N}$ controlling the relaxation of the system. If $g_0=0$, the real part of the spectrum of $\HH$ is negative and therefore the fixed point $\underline x=0$ is attractive. Increasing $g_0$, the spectrum $\rho(\l)$ of $\HH$ in the large $N$ limit consists in a flat density of complex eigenvalues contained in a circle centered at $\lambda=-1$ in the complex plane. The circle invades the positive real axes at $g_0=1$ and therefore at this point the attractor $\underline x=0$ looses stability.
Beyond this point, the dynamics is found to be confined but chaotic. At the critical point, the approach to the marginally stable fixed point is algebraic and we show that $C(t,t)\simeq {1}/{(2t)}$ when $t\to \infty$, see Fig.\ref{transition_to_chaos}.
One can also show that for $g_0<g_0^c$ and approaching the critical point, the dynamics relaxes exponentially to the fixed point $\underline x=0$ with a characteristic time that diverges as $\t \sim |g_0-g_0^c|^{-1}$.
The properties of the chaotic phase can be studied as well, following \cite{sompolinsky1988chaos}. We use as diagnostic of chaos the fact that $C(t,t')\to 0$ for $t-t'\to \infty$ and $t'\to \infty$, as we show in Fig.\ref{transition_to_chaos}. In the same figure, we also show the behavior of some individual degrees of freedom as obtained from numerical simulations, where it is clear that the dynamics is chaotic.

\subsection{Hebbian driving of synaptic plasticity}
\label{section: hebbian driving}
Eq.~\eqref{dyn_sys} describes the dynamics of a network where the interaction couplings are random and fixed in time. In \cite{clark2023theory}, Clark and Abbott considered the case in which the activity of the neurons {itself} shapes the synaptic weights, which in turn control the interaction between neurons. In our model, this is equivalent to say that the dynamics of $\underline x$ re-shapes the interaction between degrees of freedom.
In particular, following closely Clark and Abbott \cite{clark2023theory}, we consider the case where in Eq.~\eqref{dyn_sys} the current $H_i(t)$ is a function of the state of the system through 
\beq
H_i(t) = \sum_{j=1}^N A_i^j(t)x_j(t)
\label{simple_hebb_1}
\eeq
and the matrix $A_{i}^j(t)$ follows the dynamical equation
\beq
p\dot A_i^j(t) = -A_i^j(t) +\frac{k}{N}x_i(t)x_j(t)\:.\label{diff_A}
\eeq 
It is clear that the evolution of the plastic couplings $A$ depends on the overall activity of the system and the strength of $A$ depends on the coupling constant $k$, which is a control parameter.
We note that the particular form chosen for the plastic term is not mandatory.
One can easily generalize the setting to the case where
\beq
\begin{split}
H_i(t)&=\sum_{j=1}^N A_{i}^{j_1j_2\ldots j_q}(t)x_{j_1}(t)x_{j_2}(t)\ldots x_{j_q}(t)\\
p\dot A_i^{j_1j_2\ldots j_q}(t)&= -A_i^{j_1j_2\ldots j_q}(t) +\frac{k}{N^{q}}x_i(t)x_{j_1}(t)\ldots x_{j_q}(t)\:,
\label{generic_A}
\end{split}
\eeq
and for $q=1$ one gets back Eqs.~\eqref{simple_hebb_1} and \eqref{diff_A}\footnote{One could also consider the case in which Eq.~\eqref{generic_q} is replaced by a sum of terms of different order in $q$. We will not discuss this case here but this generalization is straightforward.}.
Eq.~\eqref{generic_A} can be rewitten as
\beq
A_{i}^{j_1j_2\ldots j_q}(t) = A_{i}^{j_1j_2\ldots j_q}(0) + \frac{k}{N^qp}\int_{0}^t \de s e^{-(t-s)/p} x_i(s)x_{j_1}(s)\ldots x_{j_q}(s)\:.
\label{generic_q}
\eeq
In  the following, we make the simplifying assumption that $A_{i}^{j_1j_2\ldots j_q}(0) =0$.
Inserting this form into the dynamical equation for $\underline{x}$, we get
\beq
\dot x_i(t) = -\mu(t)x_i +\xi_i(t) +\frac{k}{p}\int_0^{t}\de s e^{-(t-s)/p}C^q(t,s)x_i(s)\:.
\label{eqs_hebb}
\eeq
The DMFT equations can be easily derived from Eq~\eqref{eqs_hebb}. Using the statistical properties of $\xi_i(t)$
one gets that the dynamical system is described by an effective process given by
\beq
\dot x(t) = -\mu(t)x(t) + \xi(t) +\frac{k}{p}\int_0^{t}\de s e^{-(t-s)/p}C^q(t,s)x(s),
\label{DMFT_eff_proc_hebb}
\eeq
where
\beq
\overline \xi=0 \ \ \ \ \ \ \ \overline{\xi(t)\xi(t')} =\Xi\left[C(t,t')\right] \:.
\eeq
Multiplying Eq.~\eqref{DMFT_eff_proc_hebb} and averaging over the effective noise $\xi(t)$, we get
\beq
\partial_tC(t,t') = -\mu(t) C(t,t') +\int_0^{t'}\de s\, \Xi\left[C(t,s)\right]R(t',s) + \frac{k}{p}\int_0^{t}\de s e^{-(t-s)/p}C^q(t,s)C(t',s)\:.
\eeq
The response function $R(t,t')$ is defined as
\beq
R(t,t') = \left\langle\frac{\delta x(t)}{\delta \xi(t')}\right\rangle
\eeq
and it obeys the following dynamical equation
\beq
\partial_t R(t,t') = - \mu(t) R(t,t')+\delta(t,t') + \frac{k}{p}\int_{t'}^{t}\de s e^{-(t-s)/p}C^q(t,s)R(s,t')\:.
\eeq
At this point there are two options for the confining term $\mu(t)$.
If we impose the spherical constraint of Eq.~\eqref{spherical_con}, this implies that $C(t,t)=1$ at all times and one gets an equation for $\mu(t)$ directly by considering the equation for $C(t,t')$ and taking the limit $t'\to t$. In this way we get
\beq
\mu(t) =\int_0^{t}\de s \Xi\left[C(t,s)\right]R(t,s) + \frac{k}{p}\int_0^{t}\de s e^{-(t-s)/p}C^q(t,s)C(t,s)\:.
\eeq
If the chaotic noise is not too wild and the constraint in Eq.~\eqref{constraint_noise} holds, then we can fix $\mu(t)=f[C(t,t)]$. In this case we need to provide a dynamical equation for $C(t,t)$ which is again easily derived from the one for $C(t,t')$. We get
\beq
\begin{split}
\frac{\de C(t,t)}{\de t} &= 2\lim_{t'\to t} \partial_t C(t,t') \\
&= 2\left[\mu(t) C(t,t) +\int_0^{t}\de s\, \Xi\left[C(t,s)\right]R(t,s) + \frac{k}{p}\int_0^{t}\de s e^{-(t-s)/p}C^q(t,s)C(t,s) \right]\:.
\end{split}
\eeq
In this case, we also need to provide an initial condition for $C(0,0)=\tilde C$.
It is easy to generalize the equations when the Hebbian driving is done up to a time $t_h$ which we call the halting time, after which the coupling matrix $A_{ij}$ is fixed\footnote{One can also generalize the theory to more complex cases where the training is done with start and stop dynamics, namely when the plasticity is repeatedly switched on and off. However we do not treat this case within the DMFT but the extension is straightforward.}. Summarizing, we have the following equations for the correlation and response function
\beq
\begin{split}
\partial_tC(t,t') &= -\mu(t) C(t,t') +\int_0^{t'}\de s\, \Xi\left[C(t,s)\right]R(t',s) + \frac{k}{p}\int_0^{\tilde t}\de s e^{-(\tilde t-s)/p}C^q(t,s)C(t',s)\\
\partial_t R(t,t') &= - \mu(t) R(t,t')+\delta(t,t') + \frac{k}{p}\int_{t'}^{\tilde t}\de s e^{-(\tilde t-s)/p}C^q(t,s)R(s,t')\\
\end{split}
\eeq
and depending on whether we have a spherical or confined model we have
\begin{equation}
\begin{cases}
\mu(t) =\int_0^{t}\de s\, \Xi\left[C(t,s)\right]R(t,s) + \frac{k}{p}\int_0^{\tilde t}\de s e^{-(\tilde t-s)/p}C^q(t,s)C(t,s) & \textrm{spherical}\\
\frac{\de C(t,t)}{\de t} = 2\left[\mu(t) C(t,t) +\int_0^{t}\de s\, \Xi\left[C(t,s)\right]R(t,s) + \frac{k}{p}\int_0^{\tilde t}\de s e^{-(\tilde t-s)/p}C^q(t,s)C(t,s) \right] & \textrm{confined}
\end{cases}.
\end{equation}
The time $\tilde t$ is defined as $\tilde t=\min(t,t_h)$ and controls the dependence of the dynamics on $t_h$.
The equations above can be easily integrated numerically.
In the following, we will discuss the behavior of the solution for different values of Hebbian learning coupling $k$.

\begin{figure}
\centering
\includegraphics[width=\columnwidth]{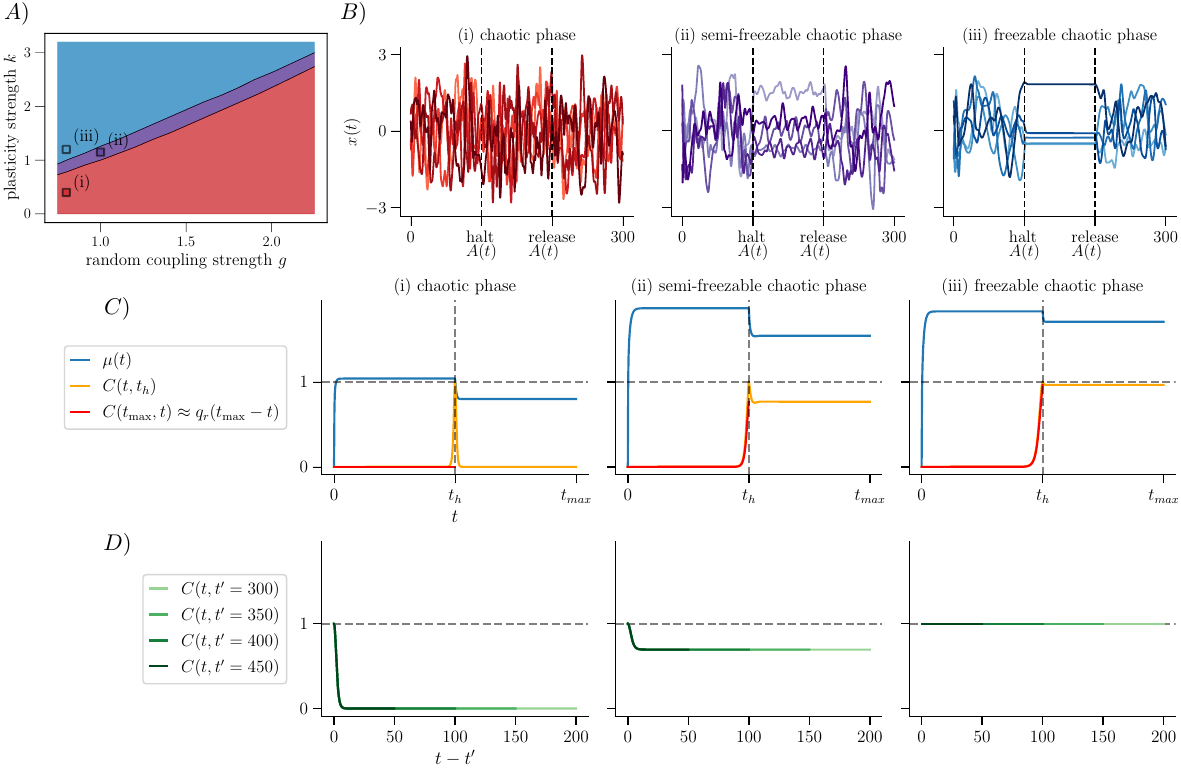}
\caption{$A$) The phase diagram of the model for $q=1$ as a function of the strength of Hebbian plasticity $k$ and of the strength of the couplings between degrees of freedom $g$. $B$) The traces of a set of randomly chosen $x_i(t)$ for a numerical simulation with $N=100$ for the plots $(i)$ and $(iii)$, and $N=200$ for plot $(ii)$. The control parameters are tuned as in the points $(i)$, $(ii)$ and $(iii)$ in the phase diagram of panel A). In the chaotic phase, halting the synaptic plasticity does not change the dynamics sufficiently enough to suppress chaos. In the semi-freezable chaotic phase instead, halting the synaptic plasticity leads the system to a chaotic attractor correlated with the configuration visited at the halting time. In the freezable chaotic phase, the dynamics converges to a fixed point attractor when the plasticity is halted. $C$) the behavior of a set of dynamical correlation functions as extracted from the numerical integration of the DMFT integrated up to time  $t_{\mathrm{max}}=500$ and fixing the halting time $t_h=250$.
In particular with the red line we plot $C(t_{\rm max},t<t_h)$ as a proxy for $q_r(t_h-t)$ and $D$) we plot with green lines $C(t,t')$ for different $t'>t_h$ and as a function of $t-t'$. This shows that for $t'$ sufficiently larger than $t_h$ the dynamics reaches a TTI regime, and that the plateau for $t-t'\to \infty$ allows to distinguish between the SFCP and the FCP.}
\label{Fig_sim_dmft_q_1}
\end{figure}

\subsection{Freezable and semi-freezable chaos}

We are now interested in the effect of plasticity on chaotic behavior. We will focus on the spherical model with 
\beq
\Xi(z) = \frac{3 g^2}{2}z^2,
\eeq
which is always chaotic for $k=0$.
Furthermore, we will consider the $q=1$ case in Eq.~\eqref{generic_q} to start with.
Following Clark and Abbott, see \cite{clark2023theory}, we consider the following protocol. 
Starting from a random initial condition on the sphere $C(0,0)=1$, we allow plastic behavior only for $t<t_h$. For $t\geq t_h$, the matrix $A_i^j(t)$ is fixed to its last value $A_i^j(t_h)$.

In \cite{clark2023theory}, Clark and Abbott have identified three phases depending on the fate of the dynamical system after the halting time $t_h$. Depending on the strength of the Hebbian learning $k$, one can distinguish three phases:
\begin{itemize}
\item \emph{Chaotic phase} (CP). At $k=0$ the system is chaotic and the halting time does not have any effect. The chaotic phase survives also when $k$ is small but finite. In this case, for $t\gg t_h$ the system completely decorrelates from the configuration at $t_h$. 
\item \emph{Semi-freezable chaotic phase} (SFCP). For an intermediate range of $k$, one observes that the dynamics is still chaotic but for $t\gg t_h$ the configurations explored are not completely decorrelated from the configuration of the system at $t_h$. Therefore, the dynamics lands on a chaotic attractor dynamically correlated with the configuration that the system had right before the halting time.
\item{\emph{Freezable chaotic phase} (FCP).} If $k$ is sufficiently large, after the halting time, the dynamics settles to a point attractor and stops. The attractor point is correlated with the configuration visited at time $t_h$. 
\end{itemize}
In order to carefully identify the tree phases, we need to consider a set of order parameters.
The two phases SFCP and FCP can be identified by looking at
\beq
q_r(0) \equiv \lim_{t\to \infty} C(t,t_h)\:.
\label{qr0}
\eeq
For both the  SFCP and FCP we have that $q_r(0)>0$, while when the system is in the CP, $q_r(0)=0$.
We can also introduce a generalization of Eq.~\eqref{qr0}. Indeed, we can consider
\beq
q_r(\Delta t) \equiv \lim_{t\to \infty} C(t,t_h-\D t)\:.
\eeq
In the FCP and in the SFCP, $q_r(\D t)$ is a positive decreasing function of $\D t$ while in the CP, we have $q_r(\D t)=0$ for all intervals $\D t$.
Therefore $q_r(\D t)$ allows to distinguish between the situation in which the system remains fully chaotic (CP) and when chaos is reduced, either completely (FCP) or not completely (SFCP).

In order to distinguish between the last two cases we need a different order parameter.
We define
\beq
q_{EA}\equiv\lim_{t,t'\to \infty, t-t'\to \infty} C(t,t')\:.
\eeq
In the FCP, we expect that $q_{EA}=1$ while $0<q_{EA}<1$ in the SFCP.
The location of the boundary between the different phases, depends on $t_h$. However, we will show that we can make some progress by looking at the asymptotic solution $t_h\to \infty$ (see Sect.\ref{section: asymptotic solution}).

The dynamical behavior in the three different phases can be visualized in the upper panel of Fig.\ref{Fig_sim_dmft_q_1}, {where we show a few traces of $x_i(t)$ for numerical simulations}. The corresponding phase diagram, as obtained from the DMFT analysis, is plotted in Fig.\ref{Fig_sim_dmft_q_1}, leftmost figure of the upper panel. All in all, the prototypical model of Eq.~\eqref{dyn_sys} under Hebbian driving displays the same phenomenology obtained in \cite{clark2023theory} with the more standard model of Eq.~\eqref{oldRNN}.

\begin{figure}[b]
\centering
\includegraphics[width=0.8\columnwidth]{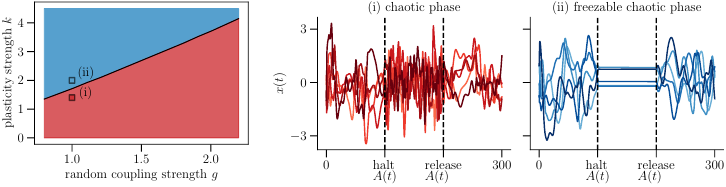}
\caption{\emph{Left panel}: The phase diagram of the spherical model for $q=2$. In this case there is no semi-freezable chaotic phase and the system undergoes a sharp transition from chaos to frozen chaos. \emph{In the middle and right panel,} we show the traces of a set of randomly chosen $x_i$ from a numerical simulation with $N=100$, in the two phases at the control parameter points denoted by $(i)$ and $(ii)$ in the phase diagram. We note that the release of plasticity in the chaotic phase accelerates the dynamics of the variables $x_i$.}
\label{fig_q_2}
\end{figure}

\subsection{The case $q=2$.}
Before looking at the asymptotic solution of the DMFT equations, we would like also to investigate the behavior of the model with $q=2$. In this case, we do not find evidence for a semi freezable chaotic phase: the system undergoes an abrupt transition from chaos to a fixed point. The corresponding phase diagram and qualitative behavior is shown in Fig.\ref{fig_q_2}.

\subsection{Asymptotic solution of the DMFT equations}
\label{section: asymptotic solution}
While the DMFT allows to explore systematically the dynamics also away from stationarity, it is useful to provide an asymptotic solution of the equations when the halting time diverges at infinity, $t_h\to \infty$.
It is clear that since the dynamics is either chaotic (fully chaotic or restricted to a sub-manifold) or it goes to a fixed point, in the asymptotic regime we expect that correlation functions become time translational invariant (TTI).
Therefore we posit, with a slight abuse of notation, that
\beq
\begin{split}
\lim_{t,t'\to \infty} C(t,t)&=C(t-t')\\
\lim_{t,t'\to \infty} R(t,t') &= R(t-t')\:.
\end{split}
\eeq
We now consider three asymptotic regimes for $t_h\to \infty$ and both $t,t'\to \infty$.

\subsubsection{\underline{Regime 1}: $t,t'\ll t_h$ with $t-t'=\D t\sim \OO(1)$.}
We first consider the regime in which $t$ and $t'$ are both diverging at infinity but they are smaller than the halting time (also diverging to infinity).
In this regime we have that plasticity is never halted for the sake of $t,t'$. Using again TTI, we consider 
the asymptotic scaling functions defined as 
\beq
\begin{split}
C_1(\D t) &= \lim_{t,t'\to \infty, t-t'=\D t} C(t,t')\\
R_1(\D t) &= \lim_{t,t'\to \infty, t-t'=\D t} R(t,t')\\
\m_{\infty}^{(1)}&=\lim_{t\to \infty, t<t_h}\mu(t)\:.
\end{split}
\eeq
Plugging this ansatz inside the dynamical equations we get
\beq
\begin{split}
\partial_{\D t}C_1(\D t) &= - \mu_\infty^{(1)}C_1(\D t)+\frac{3g^2}{2}\int_0^\infty \de s C_1^2(s+\D t)R_1(s)+\frac kp\int_{-\D t}^\infty\de s\ e^{-(\D t+ s)/p} C_1(s+\D t)C_1(s)\\
\partial_{\D t}R_1(\D t) &= - \mu_\infty^{(1)}R_1(\D t)+\frac kp\int_{-\D t}^\infty\de s\ e^{-(\D t+ s)/p} C_1(s+\D t)R_1(s)\\
\mu_\infty^{(1)} &= \frac{3g^2}{2}\int_0^\infty \de s C_1^2(s)R_1(s)+\frac kp\int_{-\D t}^\infty\de s\ e^{-s/p} C_1^2(s)\:.
\end{split}
\label{stationary_r1}
\eeq
These equations have not a causal structure since their rhs depends on times larger than $\D t$. However, they have a self-consistent structure and therefore can be solved by an iterative algorithm. One starts with a first guess of $\mu_\infty^{(1)},\  C_1$ and $R_1$ and then uses these equations to produce an updated estimate of the same quantities. We verified that this numerical procedure converges fast and is compatible with the solution of the DMFT equations, which provides a first approximation of eqs.~(\ref{stationary_r1}). 

\subsubsection{\underline{Regime 2}: $t'<t_h\ll t$ with $t_h-t'=\D t\sim \OO(1)$.}
The second asymptotic regime is obtained by considering the situation in which one of the two times $t$ and $t'$ is smaller than the halting time, while the other is larger. This regime thus controls the connection between the two stationary regimes, before and after the halting time. Since we always consider $t'<t$, we have $t'<t_h$ and $t>t_h$. Furthermore, the dynamics for $t-t_h\sim \OO(1)$ is not stationary. Therefore we consider the regime in which $t\to \infty$ and $t-t_h\to \infty$. Conversely, when $t_h-t'\sim \OO(1)$, we are probing the asymptotic stationary regime of the dynamics before the plasticity is halted and we have access to this regime thanks to Eqs.~\eqref{stationary_r1}.
In this case, the only scaling function that we have to compute is therefore
\beq
q_r(\D t) = \lim_{t,t_h\to \infty, \D t\sim \OO(1)} C(t,t_h-\D t)\:.
\eeq
Furthermore, since $t\to \infty$ we have
\beq
\mu_\infty^{(2)} =\lim_{t\to \infty, t>t_h} \mu(t)\:.
\eeq
The scaling equation for $q_r$ is found just by looking at the equations in this regime.
We get
\beq
\mu_\infty^{(2)}q_r(\D t) =  \frac{3g^2}{2}\int_0^\infty \de s q_r^2(s+\D t)R_1(s)+\frac kp\int_{-\D t}^\infty\de s\ e^{-(\D t+ s)/p} q_r(s+\D t)C_1(s)
\label{eq_qr}
\eeq
This scaling equation is not autonomous since it depends on $\mu_\infty^{(2)}$ and $C_1$ and $R_1$. The equation for $\mu_{\infty}^{(2)}$  is found by looking at the third and last asymptotic regime.

\subsubsection{\underline{Regime 3}: $t_h\ll t',t$ with $t-t'=\D t\sim \OO(1)$.}
We finally consider the last asymptotic regime in which $t,t'>t_h$ and are infinitely far from $t_h$, namely $t-t_H\to \infty$ and $t'-t_h\to \infty$.
In this case we need to consider the following scaling functions:
\beq
\begin{split}
C_2(\D t) &= \lim_{t,t'\to \infty, t-t'=\D t} C(t,t')\\
R_2(\D t) &= \lim_{t,t'\to \infty, t-t'=\D t} R(t,t')\\
\end{split}
\eeq
which obey the following scaling equations
\beq
\begin{split}
\partial_{\D t}C_2(\D t) &= - \mu_\infty^{(2)}C_2(\D t)+\frac{3g^2}{2}\int_0^\infty \de s C_2^2(s+\D t)R_2(s)+\frac kp\int_{-\D t}^\infty\de s\ e^{-(\D t+ s)/p} q_r(s+\D t)q_r(s)\\
\partial_{\D t}R_2(\D t) &= - \mu_\infty^{(2)}R_2(\D t)\\
\mu_\infty^{(2)} &= \frac{3g^2}{2}\int_0^\infty \de s C_2^2(s)R_2(s)+\frac kp\int_{-\D t}^\infty\de s\ e^{-s/p} q_r^2(s)\:.
\end{split}
\label{regime_3}
\eeq
The third scaling regime gives access to the order parameter which distinguishes between the SFCP and the FCP. Indeed we have
\beq
q_{EA} =\lim_{s\to \infty}C_2(s)
\eeq
We also note that when one is in the FCP, we have $C_2(s)=1\ \ \forall s$. 

\subsubsection{The overall structure of the asymptotic solution}
It is clear that regime 1 is fully autonomous and alone determines $C_1$, $R_1$ and $\mu_\infty^{(1)}$. Instead, we clearly see that regimes 2 and 3 are coupled by the scaling function $q_r(s)$ and by $\mu_\infty^{(2)}$. The way in which the third regime is coupled to the second is through the memory of all configurations visited for times close to $t_h$ and this is encoded in the scaling function $q_r(\D t)$. We verified that these equations are satisfied by the approximate DMFT numerical solution. 
However, we have not been able to turn Eq.~\eqref{eq_qr} into an algorithmic scheme to solve self-consistently the second and third regime. The naive iterative scheme suggested by the form of Eq.~\eqref{eq_qr} seems not convergent to the right fixed point. Nevertheless, we have checked that equations \eqref{eq_qr} and \eqref{regime_3} are coherent with the numerical solution of the DMFT equations.
All in all, this analysis shows that Hebbian driving is a powerful way to control the level of chaos in the dynamical system, as much as this happens in standard RNNs (see \cite{clark2023theory}).

\section{FORCE training}
\label{FORCE_training_sec}
Up to now, we have analyzed how a random high-dimensional chaotic system responds when Hebbian plasticity is switched on in the interactions between degrees of freedom.
However for the moment, we did not treat the case in which the dynamical system is trained to perform a task. The purpose of this section is to extend the formalism developed before to address the question of how the dynamical system can be trained to produce a desired response.

It is well known that recurrent neural networks are difficult to train by energy minimization.
Indeed, the recurrent structure of the interactions between the degrees of freedom implies that gradient signals can be indefinitely amplified due to feedback loops. Controlling this dynamics is therefore very complicated. Furthermore, it is fair to say that the extent to which one can think about biological neural network as devices that perform a gradient descent minimization is unclear \cite{whittington2019theories}. This is also because the computation of the gradient of a cost function is a complex operation that involves the so-called credit assignment problem, namely to select which control variables (or synapses) contribute the most to the error and therefore have the priority to be updated.

In order to overcome these difficulties, a number of strategies have been proposed to train recurrent neural networks. In the simplest setting, one would train a neural network such that a readout unit reproduces a complex periodic function. In other words, one sees the dynamical system as an out-of-equilibrium (chaotic) bath which generates some self-sustained dynamics and the main idea is to find a set of synaptic weights that connect the dynamical system to the readout unit so that its output is a desired one.

In this setting, one can distinguish two cases. If the readout unit is not fed back into the dynamical system, then the latter has a completely autonomous dynamics and therefore the problem of the explosion of gradients in a putative energy minimization training dynamics is mostly solved. This idea has been exploited enormously in the past and it is at the basis of Echo-state or Liquid-state networks \cite{jaeger2001echo, maass2002real, jaeger2004harnessing}.

A more complex setting consists in the situation where the output of the readout unit is re-injected into the dynamical system itself. 
This setting can be seen as a simplified version of training a single neuron and leaving the rest of the network unaltered. Given that the output of the readout neuron is fed back into the network, this setting suffers of the same instabilities of more general recurrent neural networks.
In 2009, Sussillo and Abbott \cite{sussillo2009generating} have shown that one can efficiently train the readout unit coupled to the dynamical system in Eq.~\eqref{oldRNN} via an algorithmic strategy called FORCE, which stands for First-Order Reduced and Controlled Error. The main idea of the algorithm is that the synaptic weights are updated always by keeping the error small along the whole training dynamics.
The algorithm can be extended in many more complex situations, and more recently, it has been also shown that one can use it to train a set of Spiking Neural Networks (SNNs) \cite{nicola2017supervised} which differ from Eq.~(\ref{oldRNN}) because the dynamics of the membrane potential is resolved in time and the rates are computed microscopically as the number of times an action potential is fired.

It is fair to say that while numerical simulations have shown that FORCE can train recurrent neural networks with thousands of neurons, it is anyway unclear how the algorithm behaves on instances of infinite system size. This may be important for large scale neural networks and in particular for biological ones. The purpose of this section is to explore the performance of the FORCE algorithm
in the context of the high-dimensional chaotic systems of the form represented in Eq.~\eqref{dyn_sys}
and to construct a mean field theory analysis of such algorithms.

\subsection{FORCE algorithm}\label{sec_FORCE}
We will first recall here the setting and the algorithm introduced in \cite{sussillo2009generating} and then we will adapt it to our setting.
We first consider the Eq.~\eqref{oldRNN} and introduce an input current of the form
\beq
H_i(t) = w_i^{(f)}z(t) \:.
\eeq
The variable $z(t)$ is the output of the readout unit. In the simplest setting we consider 
\beq
z(t) = \sum_{i=1}^N w_i^{(o)} r_i(t),
\eeq
so that the output unit performs a linear readout of the state of the system.
We have two sets of weights: $w_i^{(o)}$ are the synaptic weights connecting the dynamical system to the readout unit and these are the variables that we want to change in order to perform a task.
The weights $w_i^{(f)}$ are instead the feedback weights and are supposed to be fixed.
The taks we want the network to learn is to reproduce a function. Consider a periodic function $f(t)$ with period $T$. We would like to find that at the end of the training phase, the output of the readout neuron is $z(t)=f(t)$.
In this way, learning will correspond to turn the chaotic noise of the dynamical activity of the untrained network to a more structured response.
This task is the simplest one that cannot be performed without a feedback of the output neuron into the network itself\footnote{Simpler tasks like classification can instead be performed without feedback from the readout unit.}.
In \cite{sussillo2009generating}, Sussillo and Abbott have proposed the following training strategies to find a good set of weights $w_i^{(o)}$. In order to define them properly, we assume that the dynamical system in Eq.~\eqref{oldRNN} is discretize with time step $\de t$. Then we can define two algorithms:
\begin{enumerate}
\item \underline{FORCE-I} \cite{sussillo2009generating}: In this case we first define
\beq
z^+(t) = \underline w^{(o)}(t) \cdot \underline r(t+\de t)
\eeq
and we update
\beq
\underline w^{(o)}(t+\de t) = \underline w^{(o)}(t) -\eta(t+\de t) \left(z^+(t)-f(t+\de t)\right)\underline r(t+\de t)\:.
\eeq
Therefore in order to run the dynamics, in this case one first needs to update the dynamical variables $\underline r(t)$ and then the weights $\underline w(t)$. The learning rate $\eta(t)$ is a control parameter of the problem.  It is known that this algorithm, while being more biologically plausible,  suffers from instabilities and can learn only simple tasks \cite{sussillo2009learning}. These problems have been solved numerically by developing a different, more complex, and less biologically plausible algorithm, which is FORCE-II.
\item \underline{FORCE-II} \cite{sussillo2009generating}: The update rule for the output weights is different. We define the error
\beq
e_-(t) = z^+(t-\dt)-f(t)
\eeq
and update the weights with the following scheme
\beq
\underline w^{(o)}(t+\de t) = \underline w^{(o)}(t) - e_-(t+\de t) P(t+\de t)\underline r(t+\de t)\:.
\eeq
The matrix $P(t)$ is an $N\times N$ matrix which follows a dynamical evolution given by the update rule
\beq
\begin{split}
P(0) &= \frac{1}{\alpha} \mathbf{1}  \\ 
P(t+\de t) &= P(t) - \frac{P(t)\underline r(t) \underline r(t)^T P(t)}{1+\underline r(t+\de t)^TP(t)\underline r(t+\de t)}
\end{split}
\eeq
and we have indicated by $\mathbf 1$ the identity matrix. The parameter $\alpha$ is a control parameter of the algorithm. This algorithm is naturally formulated in discrete time.
\end{enumerate}

We now adapt both algorithms to train the dynamical system in Eq.~\eqref{dyn_sys}. 
In order to simplify the formalism, we first consider $w_i^{(f)}=1$ for all $i=1,\ldots,N$. We underline that the formalism we are going to develop can be generalized to the case in which $w_i^{(f)}$ is taken to be random.
In this way, we have only the set of weights that define the output unit and we call them $\underline w$.
Therefore we define
\beq
z(t) = \frac 1N \underline w(t) \cdot \underline x(t) 
\eeq
and we assume that the task of the learning protocol is to get $z(t)=f(t)$ at the end of learning.
Both FORCE algorithms are formulated in terms of the variables $\underline x$ and $\underline r$. However, the dynamical system in Eq.~\eqref{dyn_sys} has only the $\underline x$ as degrees of freedom. In order to take into account this and the $N\to \infty$ limit, we consider a modified version of FORCE adapted to our setting.
\begin{itemize}
\item \underline{FORCE-I}: we define
\beq
z^+(t) = \frac 1N \underline w (t) \cdot \underline x(t+\de t)
\label{def_zplus}
\eeq
and we update the weights according to 
\beq
\underline w(t+\de t) = \underline w(t) -\eta(t+\de t) \left(z^+(t)-f(t+\de t)\right)\underline x(t+\de t)
\label{forceIupdate}
\eeq
It is very easy to show that at each time step, this algorithm is built in such a way that $f(t)=z(t)$ if $\eta(t)$ is carefully chosen (see below).
\item \underline{FORCE-II}: also in this case we define
\beq
e_-(t) = z^+(t-\dt)-f(t)
\eeq
and we update the weights according to
\beq
\underline w(t+\de t) = \underline w(t) - e_-(t+\de t) P(t+\de t)\underline x(t+\de t)\:.
\label{update_force_II}
\eeq
The matrix $P(t)$ follows the dynamical evolution
\beq
\begin{split}
P(0) &= \frac{1}{\alpha} \mathbf{1}  \\ 
P(t+\de t) &= P(t) - \frac 1N \frac{P(t)\underline x(t+\dt) \underline x(t+\dt)^T P(t)}{1+\frac 1N\underline x(t+\de t)^TP(t)\underline x(t+\de t)}
\end{split}.
\label{update_P_rescaled}
\eeq
This algorithm works by keeping the error, namely $z(t)-f(t)$, small as time increases. In particular, we will study how the error decreases during learning.
\end{itemize}

\subsection{Numerical simulations}

In this section, we present a set of numerical simulations to show that the FORCE algorithm –as detailed in Sec.~\ref{sec_FORCE} and adapted to a random dynamical system– works to train it efficiently. We will focus on FORCE-II since FORCE-I can only be used to train simple functions \cite{sussillo2009generating}.
We consider the confined model in Eq.~\eqref{dyn_sys} with $\mu(t)=C(t,t)$ and integrate numerically the dynamical equations at fixed learning rate equal to $\de t=0.01$. In particular, we consider $N=100$, $\alpha=0.001$ and $\hat{g}=0.7\sqrt{3/4}$ for all the data-set plotted in this section.

\begin{figure}
\centering
\includegraphics[width=\columnwidth]{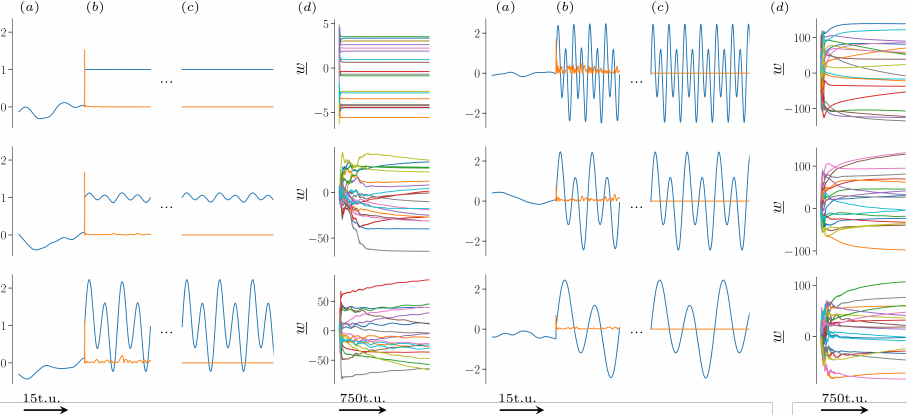}
\caption{The performance of FORCE-II to train the confined model. \emph{Left panel}: Performances when training a periodic perturbation of varying amplitude around a constant. The \emph{blue curve} is the network output $z(t)$, while the \emph{orange curve} is the norm of the weight vector variation $\de \underline w = \underline w(t+\de t) - \underline w(t)$. \emph{(a)}: Before training: Chaotic dynamics. The weights $w_i$ are randomly uniformly initialised in $[-5,5]$ so the network output $z(t)$ oscillates around $0$. \emph{(b)}: Training phase. FORCE-II drives quickly the output $z(t)$ to generate the target one so the weight update $\|\de \underline w\|$ is big initially and then decreases. In the left panel, $\|\de \underline w\|$ has been re-scaled by a factor $0.1$ to allow the plot to fit better into the figure. Only the beginning of training is plotted. \emph{(c)}: After training. Once $\|\de \underline w\|$ is sufficiently small (ideally of order $10^{-5}$), training can be switched off. The output weight $\underline{w}$ is then fixed to its last value during training and $z(t)$ autonomously produces the target output (which is not plotted given that it superimposes to $z(t)$). Instead the plot of $\|\de \underline w\|$ (orange curve) is constant equal to $0$ since $\underline w$ is fixed. \emph{(d)}: Evolution of $\underline{w}$ during the training phase. Training lasted in total 1500 time units (t.u.) in all the plots to allow for comparison. \emph{Right panel:} Same analysis as in the left panel but with the target being a periodic function with varying frequency.} 
\label{sim_around_const}
\end{figure}

In Fig.~\ref{sim_around_const} left panel, we consider the task of learning first a constant output $f(t)=1$ and progressively add a small periodic perturbation around the constant value. Specifically, we choose $f(t)=A+B g(t,\underline \omega^*)$ with $A=1$, and $B$ changes on each row of the figure. In the first row we have $B=0$, while for the second row $B=0.1$, and the last $B=1$.
The function $g(t,\underline \omega)$ is defined as
\beq
\begin{split}
g(t,\underline \omega)&=\frac{1}{\sqrt{a^2+b^2}}\left(a \sin(2\pi \omega_0 t) + b\sin(2\pi \omega_1 t)\right)\\
\underline \omega^*&=\{\omega_0=0.1,\ \omega_1=0.2\}\\
a&=0.6\\
b&=1.2\:.
\end{split}
\eeq 
In the right panel of Fig.\ref{sim_around_const}, we play the same game as learning a periodic function, this time changing the frequency and without any constant offset. In particular, we learn $f(t)=2g(t,y \underline \omega^*)$ with $y=0.5, 1, 2$ from bottom to top.

If the strength of the chaotic term is not too large (see Sect. \ref{section: numerical integration of DMFT: performance} for a precise way to quantify how strong it can be), we see that learning is possible. In this case when learning is switched on, the output of the network almost instantly matches the target function, which is a necessary condition for a successful FORCE-training  \cite{sussillo2009generating}. As learning proceeds, the readout weights $w_i$ should reach time-independent values. In practice however, we observe that reaching $\|\underline w(t+\de t) - \underline w(t)\| \approx 10^{-4}$ at the end of training gives satisfying performances in the testing phase after training. We also note that the amplitude and frequency of the target function influence the training process. In the left panel of Fig.\ref{sim_around_const}, the larger the amplitude $B$ of the periodic perturbation, the slower learning takes place; while in the right panel of Fig.\ref{sim_around_const}, a periodic function with an intermediate frequency characterized by $y=1$ is learned faster than one with $y=0.5$ or $2$.

\begin{figure}[b]
\centering
\includegraphics[width=1\columnwidth]{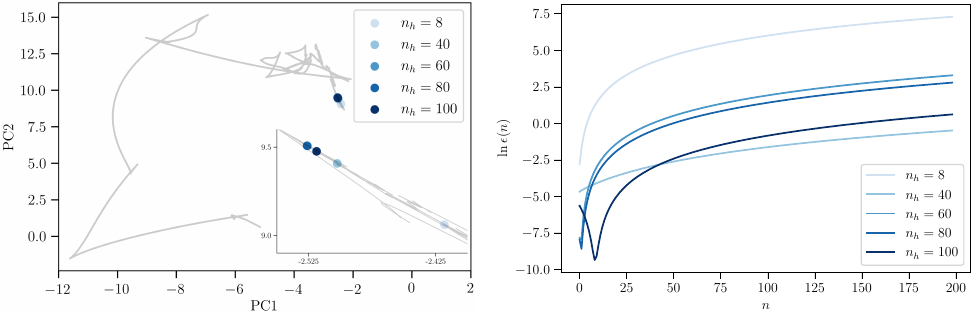}
\caption{\emph{Left Panel}:  {Trajectory of the weight vector $\underline w$ (\emph{gray} trace) during one learning episode, projected on the first two principal components (PCs) of the dynamically averaged correlation matrix $\langle \underline{x}^T(t)\underline{x}(t) \rangle_t$} (explained variance of the first two PCs, $0.978$). Each component $w_i$ is initialized randomly from a uniform distribution in $[-15,15]$. As training proceeds, the dynamics reaches quickly a small region of phase space yielding good performances. The \emph{blue} dots indicate the state of the system after learning for $n_h = 8,\,40,\,60,\,80,\,100$ periods of the target function $f(t)$. The \emph{indent} figure is a zoom on the end of the trajectory, showing mild fluctuations even after learning has converged satisfyingly. Note that the dots for $n_h=40$ and $n_h=60$ are superimposed. \emph{Right Panel}: The error after training as a function of time, measured in the number $n$ of periods of $f(t)$. The \emph{blue} curves show the errors after training for different periods of $f(t)$. If the training time is not large enough, soon after training the dynamics is not able to stay close to the desired output. Instead for larger training times, one reaches a configuration where the performance fluctuates also due to finite size.}\label{PCA}
\end{figure}

Fig.\ref{PCA} instead sheds light on the small region of phase space reached by the dynamical system during training and shows how stable that region is after training, as a function of training time. We consider one learning episode during which the output $z(t)$ is trained to reproduce the target $f(t)=3\sin (t/2)/2$. In the left panel, we plot the projection of $\underline w(t)$ during training on the first two principal components (PCs) of the auto-correlation matrix $\langle \underline{x}^T(t)\underline{x}(t) \rangle_t$, which is computed once $f(t)$ has been learned. The dots on this plot represent the position in the projected PC space of the dynamics after 8, 40, 60, 80 and 100 periods of $f(t)$. Thus, we see that the dynamics converges very fast to a small region of phase space where $z(t)$ matches $f(t)$, and then moves very slowly in that region. In the right panel of the same figure, we also plot the performance of the network if –during the same learning episode– we stop training after $n_h=$ 8, 40, 60, 80 and 100 periods of $f(t)$. The performance of the network is measured with the error $\epsilon(n)$, defined as the temporal average of the squared difference between $f(t)$ and $z(t)$ evaluated along the $n$-the period of the target function $f(t)$
\beq
\epsilon(n)=\int_{nT}^{(n+1)T}\de s |f(s)-z(s)|^2\:.
\eeq
In the right panel of Fig. \ref{PCA}, we see that the earlier we stop learning, the worst the performances. But after a while $\left( n_h \geq 40\right)$, performance fluctuates as the dynamics wanders in the small region of phase space yielding small errors.
All in all, Fig. \ref{sim_around_const} shows that the dynamical system in Eq.~\eqref{dyn_sys} can be trained with the FORCE-II algorithm that we have described in Sec.~\eqref{sec_FORCE}, see Eqs.~\eqref{def_zplus}-\eqref{update_P_rescaled}.

\subsection{Dynamical mean field theory of FORCE training}
In recent years, there has been a growing interest in trying to apply DMFT to study learning in ANNs, especially in supervised learning settings with feed-forward networks \cite{mignacco2020dynamical, mignacco2021stochasticity, bordelon2022self, kamali2023stochastic}.
In this section, we develop a DMFT analysis of both FORCE algorithms, which to the best of our knowledge has not been performed before.
Since both algorithms are defined in the discrete time setting, we use the confined model for the dynamical system in order to follow its trajectory exactly in the large $N$ limit.

\subsubsection{The DMFT equations for the dynamical system}
We first describe the DMFT for the dynamical system in Eq.~\eqref{dyn_sys} when the input current is given by $H_i(t)=z(t)$. We assume that time is discretized by a time step $\de t$.
Using the same arguments as before, one can show that the DMFT equations are
\beq
\begin{split}
C(t+\de t, t') - C(t,t') &= \de t \left[-C(t,t) C(t,t') + \frac{3 g^2}{2}\sum_{i=0}^{t'/\de t}C^2(t,i \de t) R(t',i\de t) + z(t)m(t')\right] \\
C(t+\de t, t+\de t) - C(t,t) &= 2\de t \left[-C(t,t)^2 + \frac{3 g^2}{2}\sum_{i=0}^{t/\de t}C^2(t,i \de t) R(t,i\de t) + z(t)m(t)-\de t C(t,t)z(t)m(t)\right]\\
&+\de t^2\left[\frac{3g^2}{2}C^2(t,t)+C^3(t,t)+z^2(t)-3g^2C(t,t)\sum_{i=0}^{t/\de t}C^2(t,i \de t) R(t,i\de t)\right]\\
R(t+\de t,t') -R(t,t')&= - \mu(t) R(t,t')\de t + \delta_{t/\de t, t'/\de t}\\
m(t+\de t) - m(t) &= \de t \left[- \mu(t) m(t) + z(t)\right]\\
\mathrm{with} \; \;
C(0,0)&=\tilde C\\
R(0,0)&=m(0)=z(0)=0\:.
\end{split}
\label{dyn_FORCE}
\eeq
The function $m(t)$ controls the magnetization of the system and it corresponds to
\beq
m(t) = \frac 1N \sum_{i=1}^N x_i(t)\:.
\eeq
In the large $N$ limit, $m(t)$ concentrates on its average (over the initial conditions of the dynamics and over the random realization of the chaotic noise term).
Finally, the initial conditions for the dynamical correlators are due to the fact that we assume that 
the initial condition for $x_i(0)$ is extracted from a Gaussian measure with variance $C_d$ and that $\underline w(0)$ is a vector with zero mean and uncorrelated with $\underline x(0)$. In all our numerical integration we considered $\tilde C=1$.

From the point of view of the dynamical system of the $\underline x$ variables, the dynamics of the output unit is fully encoded in the variable $z(t)$. Therefore, the rest of the DMFT analysis concerns the characterization of the dynamical evolution of $z(t)$.
Since we have two FORCE algorithms, we will now describe their corresponding DMFTs.

\subsubsection{DMFT of FORCE-I}
We need to consider both Eq.~\eqref{def_zplus} and Eq.~\eqref{forceIupdate}. Eq.~\eqref{def_zplus} defines a scalar function, $z^+(t)$ which concentrates in the high dimensional limit. The goal of the DMFT analysis is to provide an equation for $z(t)$ and $z^+(t)$. Using Eq.~\eqref{forceIupdate} it is easy to show that
\beq
z(t+\de t) = z^+(t) - \eta(t+\de t) (z^+(t)-f(t+\de t))C(t+\dt,t+\dt) 
\label{zplus_dmft}
\eeq
In particular, this implies that if we choose $\eta(t+\de t) = 1/C(t+\dt,t+\dt)$ the dynamics during training runs on an error free trajectory since $z(t)=f(t)$ at all times.
In order to close the DMFT analysis, we need to provide an equation for $z^+(t)$. This can be obtained by noting that the equation for $\underline w(t)$ can be rewritten as
\beq
\underline w(t) = \underline w(0)-\sum_{i=0}^{(t-\dt)/\dt}  \eta((i+1)\dt) \left(z^+(i\dt)-f((i+1)\de t)\right)\underline x((i+1)\dt)\ \ \ \ \ \ \ t\geq \dt
\eeq
Therefore, if we assume that $\underline w(0)=0$, it is easy to show that
\beq
z^+(t) = -\sum_{i=0}^{(t-\dt)/\dt}  \eta((i+1)\dt) \left(z^+(i\dt)-f((i+1)\de t)\right)C(t+\dt,(i+1)\dt)\ \ \ \ \ \ \ t\geq \dt\:.
\label{z_dmft_force_1}
\eeq
So Eqs.~\eqref{dyn_FORCE},  Eq.~\eqref{zplus_dmft} and Eq.~\eqref{z_dmft_force_1} define a causal system of equations that can be integrated numerically. They describe the behavior of the FORCE-I algorithm in the $N\to \infty$ limit.
It is interesting to see that the behavior of the function $z^+(t)$ depends on $C(t,t')$ and therefore somehow has a memory of the system's history.

\subsubsection{DMFT of FORCE-II}
This case is more complicated due to the fact that the dynamics of the weights of the output unit depends on the dynamics of the matrix $P$ which has a more complex flow equation.
However, we will show that this change can be anyway taken into account in the high-dimensional limit.
First of all, we consider Eq.~\eqref{update_force_II} and multiply it by $\underline x(t+\dt)/N$. We get
\beq
z(t+\dt) = z^+(t)-e_-(t+\dt)\PP(t+\dt,t+\dt,t+\dt),
\label{z_II}
\eeq
where we have denoted
\beq
\PP(t,t',t'')=\frac 1N \underline x(t)^TP(t')\underline x(t'')\:.
\eeq
Using the same argument as for FORCE-I, we can also write
\beq
z^+(t)= -\sum_{i=1}^{t/\dt}  e_-(i\dt) \PP(t+\dt,i\dt, i\dt)\ \ \ \ \ \ \ t\geq \dt\:.
\label{zplus_II}
\eeq
It is clear that the exact solubility of the DMFT relies on the ability to find a recursion relation for the matrix elements of the operators $P(t)$.
We will now show that such matrix elements can be obtained by recursive relations in terms of the correlation functions $C(t,t')$.
First of all we have that
\beq
\PP(t,0,t')=\frac 1\alpha C(t,t')\:.
\eeq
Furthermore, the dynamical equation for $P$ gives
\beq
\PP(t,s+\dt,t') = \PP(t,s,t')-\frac{\PP(t,s,s+\dt)\PP(s+\dt,s,t')}{1+\PP(s+\dt,s,s+\dt)}\:.
\eeq
It is easy to convince oneself that this system of equations has a causal structure and therefore can be integrated numerically very easily.
Therefore together with Eq.~\eqref{dyn_FORCE}, we have the full DMFT equations given by
\beq
\begin{split}
&z(t+\dt) = z^+(t)-e_-(t+\dt)\PP(t+\dt,t+\dt,t+\dt)\\
&z^+(t) = -\sum_{i=1}^{t/\dt}  e_-(i\dt) \PP(t+\dt,i\dt, i\dt)\\
&\begin{cases}
\PP(t,0,t')&=\frac 1\alpha C(t,t')\\
\PP(t,s+\dt,t') &= \PP(t,s,t')-\frac{\PP(t,s,s+\dt)\PP(s+\dt,s,t')}{1+\PP(s+\dt,s,s+\dt)}\:.
\end{cases}
\end{split}
\label{P_eq_II}
\eeq 
It is also clear that we can consider a continuous time limit leading to partial differential equations.
We do not investigate this point in this work.

\subsubsection{Generalizations}
It is also useful to generalize the formalism presented above to the case in which there are $k$ output neurons performing a linear readout of the system.
In this case we consider that we have the input currents in the dynamical system given by
\beq
H_i(t) = \sum_{l=1}^k c_l z_l(t)
\eeq
where $c_l$ are constants that are fixed and of order one. 
We denote by $z_l(t)$ the output of the $l$ unit and
\beq
z_l(t) = \frac 1N \underline w_l\cdot \underline x(t)
\eeq
where $\underline w_l$ are the weights of the $l$ readout unit.
We assume that there is no connection between the linear readout units and that they interact only via their feedback loops onto the dynamical system. In this case, the task would be that each readout unit produces a target function $f_l(t)$ for $l=1,\ldots k$.  
It is clear that the DMFT equations for the dynamical system can be straightforwardly generalized. We get
\beq
\begin{split}
C(t+\de t, t') - C(t,t') &= \de t \left[-C(t,t) C(t,t') + \frac{3 g^2}{2}\sum_{i=0}^{t'/\de t}C^2(t,i \de t) R(t',i\de t) + m(t')\sum_{l=1}^kc_l z_l(t)\right] \\
C(t+\de t, t+\de t) - C(t,t) &= 2\de t \left[-C(t,t)^2 + \frac{3 g^2}{2}\sum_{i=0}^{t/\de t}C^2(t,i \de t) R(t,i\de t) +m(t)\sum_{l=1}^kc_l z_l(t)\right]\\
&+\de t^2\left[\frac{3g^2}{2}C^2(t,t)+C^3(t,t)+\left(\sum_{l=1}^kz_l(t)\right)^2-3g^2C(t,t)\sum_{i=0}^{t/\de t}C^2(t,i \de t) R(t,i\de t)\right]\\
R(t+\de t,t') -R(t,t')&= - \mu(t) R(t,t')\de t + \delta_{t/\de t, t'/\de t}\\
m(t+\de t) - m(t) &= \de t \left[- \mu(t) m(t) + \sum_{l=1}^kc_l z_l(t)\right]\\
\mathrm{with} \; \;
C(0,0)&=\tilde C\\
R(0,0)&=m(0)=z_l(0)=0\:.
\end{split}
\eeq
Since we know that there is no direct interaction between the readout units, it is easy to perform the FORCE algorithm on all of them. We focus on FORCE-II. It is easy to show that for each $l=1,\ldots k$ we have a generalization of the DMFT equations for FORCE-II given by
\beq
\begin{split}
&z_l(t+\dt) = z^+_l(t)-e_-(t+\dt)\PP_l(t+\dt,t+\dt,t+\dt)\\
&z^+_l(t) = -\sum_{i=1}^{t/\dt}  e_-(i\dt) \PP_l(t+\dt,i\dt, i\dt)\\
&\begin{cases}
\PP_l(t,0,t')&=\frac 1\alpha C(t,t')\\
\PP_l(t,s+\dt,t') &= \PP_l(t,s,t')-\frac{\PP_l(t,s,s+\dt)\PP_l(s+\dt,s,t')}{1+\PP_l(s+\dt,s,s+\dt)}\:.
\end{cases}
\end{split}
\eeq
It is clear that if $c_l=c$ and $f_l(t)=f(t)$ for all $l=1,\ldots, k$ the system has a mode collapse where all output neurons become the same. An interesting question would be how does the system behaves as soon as there is some small deviation from this rather symmetric situation. Can we understand the solution of the DMFT in terms of perturbation theory? This is left for future work. 
We note that the integration of the DMFT equations in this case is highly parallelizable. Indeed, each output neuron runs independently of the other and the only inputs needed are the dynamical correlation functions $C(t,t')$.

\subsubsection{Numerical integration of the DMFT dynamics: performance of the algorithms}
\label{section: numerical integration of DMFT: performance}
In this section we show the results of the numerical integration of the DMFT equations describing FORCE-II, see Eqs.~\eqref{dyn_FORCE} and \eqref{z_II}-\eqref{P_eq_II}.
We separate two cases, a simple case where the network needs to learn a constant function and the case in which it has to learn a periodic function. {In all numerical integration we work with $dt=0.1$ and $\alpha=0.001$.}

\begin{figure}
\centering
\includegraphics[width=0.8\columnwidth]{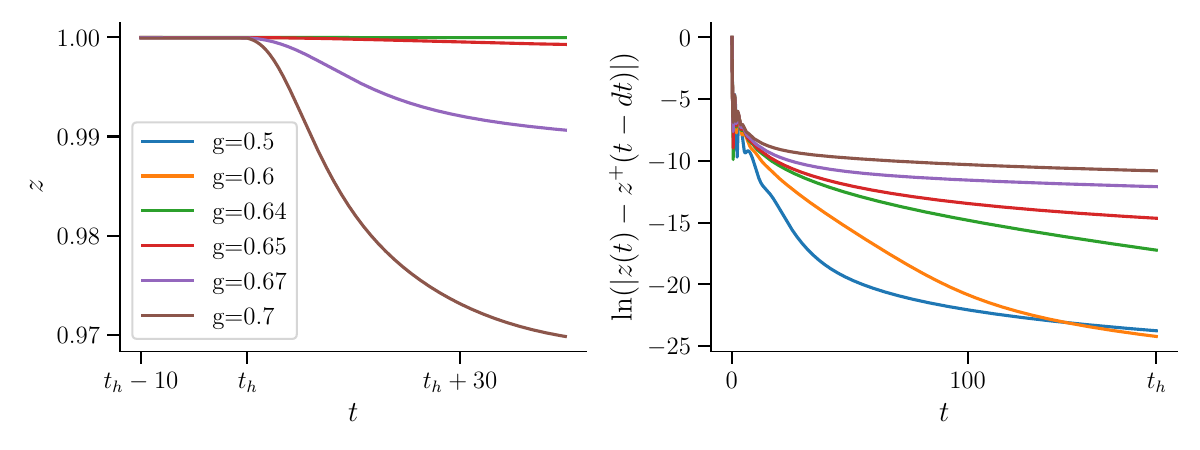}
\caption{\emph{Left Panel}: the output of the network is trained to reproduce a constant function {for times $t<t_h$. After $t_h$, training is stopped.} If the chaos level is sufficiently small the network can be trained to stay close to the target output. \emph{Right Panel}: the difference of $z(t)-z^+(t-\de t)$ as a function of time for different values of $g$. {If $g$ is larger than $0.64$, the difference $z(t)-z^+(t-\de t)$ stays large and the network cannot be successfully trained.}}
\label{constant_FII_DMFT}
\end{figure}

\paragraph*{Learning a constant function --}
We consider the dynamical system trained with FORCE-II to reproduce a constant function $f(t)=1$.
In the left panel of Fig.\ref{constant_FII_DMFT} we plot the output of the network $z(t)$ as a function of time across the end of the training phase and at the beginning of the post-training phase, for different values of the coupling constant $g$ tuning the strength of the chaotic noise term.
We clearly see that as soon as $g$ is smaller than a critical value which is reasonably estimated between $0.64$ and $0.65$, the post-training phase is good and the system has been able to go to a fixed point. Conversely, if chaos is too strong the network is not able to stay close to the constant output. In the right panel of the same figure we plot the difference between the output $z(t)$ and $z^+(t-\de t)$. This difference is actually proportional to $\de \underline w(t) = \underline w(t)-\underline w(t-\de t)$ and therefore, if it decays to zero, it means that $\|\de \underline{w}\| \to 0$ and the output unit is reaching a fixed point. For small values of $g$ it seems that this is the case, while for larger values of $g$, the output is not converging to a fixed point.
In order to understand the critical value of $g$ at which learning becomes possible, we can easily argue as follows. FORCE-II drives the dynamical system to $f(t)=1\equiv f_0$ across the training phase.
If this drive is sufficient to let the dynamical system approach a fixed point, then the post-training phase will be such that the system stays at the attractor induced by the constant force $z(t)=f_0$.
Therefore, the phase diagram can be drawn by looking at whether a constant force $z(t)=f(t)=f_0$ is sufficient to suppress chaos and induce an attractor in the dynamical system. This will be possible only if the level of chaos is sufficiently small with respect to $f_0$.

To understand the critical chaos strength, we assume that for $z(t)=f_0$ the dynamical system goes to a fixed point. The equations describing the fixed point are easily derived from the statistical properties of the chaotic noise term. Denoting
\beq
C_d= \lim_{t\to \infty} C(t,t)
\eeq 
we get that
\beq
C_d=\frac{1}{C_d^2} \left(\frac{3g^2}2 C_d^2 + f_0^2\right)\:.
\label{Cd_fixed_point}
\eeq
In order to understand if this equation describes a fixed point, we need to compute its stability. Let us denote the coordinates of the fixed point as $\underline x^{(0)}$. Assuming that $\de t\to 0$ and expanding the dynamical system around this point, $x_i=x_i^{(0)}+\delta_i$ we get
\beq
\dot \d_i = -\sum_{j=1}^N M_{ij} \d_j
\eeq
The stability of the fixed point is controlled by the real part of the spectrum of $M$. 
The matrix $M$ is given by
\beq
M_{ij} = C_d\d_{ij} +\frac 2N x_i^{(0)}x_j^{(0)} + \frac{2\hat g}{N} \sum_{k=1}^N J^i_{jk} x_k^{(0)}\:.
\eeq
It is easy to show that the real part of the spectrum of this random matrix touches zero when\footnote{Note that the matrix $M_{ij}$ contains a low rank projector. However, depending on $g$ this term may give rise to an isolated eigenvalue on the right of the bulk of the spectrum, and since here we are mostly focusing on the left side of the spectrum, this term is harmless.}
\beq
C_d=2\hat g\sqrt{C_d}
\eeq
and therefore learning can take place only for 
\beq
g<g_c=\sqrt{\frac{C_d}{3}}\:.
\eeq
Using Eq.~\eqref{Cd_fixed_point} we get
\beq
g_c = \frac{1}{\sqrt 3} \left(2f_0^2\right)^{1/6}\:.
\eeq
Therefore if $g<g_c(f_0)$, the dynamical system can learn a constant function $f_0$. If $f_0=1$ we get $g_c\simeq 0.648$ which agrees with the numerical integration of the DMFT equations (see Fig. \ref{constant_FII_DMFT}).

\paragraph*{Learning a periodic function --}

\begin{figure}[h]
\includegraphics[scale=0.7]{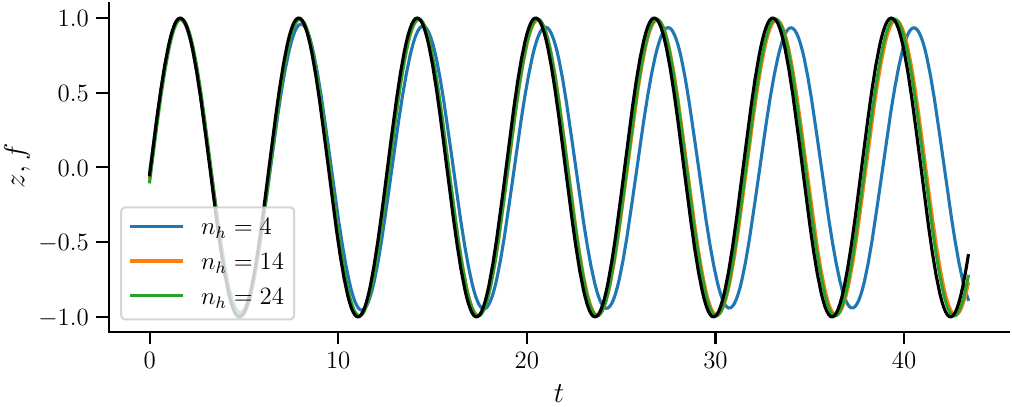}
\caption{The output $z(t)$ as obtained from the numerical integration of the DMFT equations, as a function of time in the post-training phase. The black line is the target output function $f(t)$. In lighter colors, the output for different values of the total learning time measured in number of periods of the function $f(t)$.}
\label{out_ex}
\end{figure}

\begin{figure}[h]
\centering
\includegraphics[width=0.8\columnwidth]{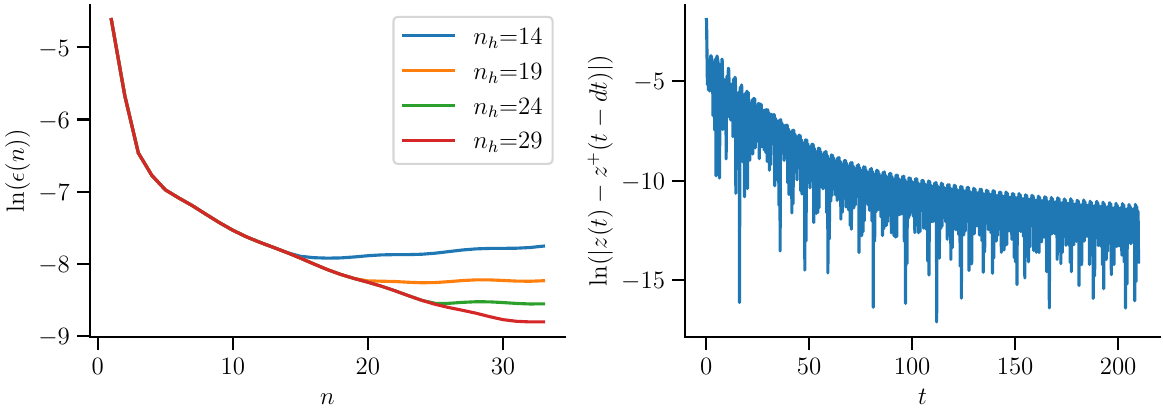}
\caption{\emph{Right Panel}: the error during training and post training for different values of the training time measured in the number of periods of $f(t)$ for $f(t)=3\sin(t)/2$. \emph{Left Panel}: the difference between $z(t)$ and $z^+(t-\de t)$ which confirms that during training the dynamics is converging to a fixed point.}
\label{performance_fig}
\end{figure}

In Fig.\ref{out_ex}, we plot the output of the network in the post-training phase, as obtained by numerically integrating the DMFT equations, when the network is trained with different training times (measured in terms of periods of the periodic function $f(t)$). We choose to train the network on a simple sinusoidal function. We clearly see that the output stays closer to the black line (the function $f(t)$) the larger the number of training periods.

In order to better characterize this behavior, in the right panel of Fig.\ref{performance_fig} we plot the error $\epsilon(n)$ as a function of training periods, for different values of the training time measured in the number of periods of the function $f(t)$. We see that as soon as the training stops,
the error increases exponentially, albeit with a rate that is smaller the larger the training time. Furthermore in the right panel, we plot the difference $z(t)-z^+(t-dt)\propto (\underline w(t)-\underline w(t-\de t))$ during training. We clearly see that FORCE-II is exponentially converging to an attractor and therefore this algorithm is effective in training the dynamical system in the infinite system size limit.

\section{Conclusion and perspectives}
\label{perspectives_sec}
We considered a simple set of high-dimensional chaotic systems and compared their dynamical behavior to standard RNNs under various driving forces and mechanisms. We showed in Sec. \ref{Hebbian_training_sec} that this class of models has chaotic properties and phases analogous to what was shown by Sompolinsky et al. \cite{sompolinsky1988chaos} and Clark and Abbott \cite{clark2023theory} in more standard models of RNNs, thus establishing these models as good abstract models of more biologically grounded RNNs. We then showed in Sec. \ref{FORCE_training_sec} that the prototypical models we analyzed could also be trained via the FORCE algorithm to generate simple periodic patterns and we believe that this opens the way to study in detail the learning dynamics of more standard RNNs.

We now list a number of possible extensions of our approach, which can be studied using the methods developed in this work.
\begin{enumerate}
\item {\emph{The phase space of the readout weights $\underline w$}.--} The DMFT analysis of FORCE can be simply closed on the dynamics of the scalars $z(t)$ and $z^+(t)$. However it would be very interesting to understand the dynamics of the weights $\underline w$. This is accessible from our formalism but we leave a detailed investigation for future work. Looking at this would clarify what is the feasible phase space of the linear readout vectors and how this space is explored by the learning algorithms. A complementary question is also related to the complexity of the function the system needs to learn. While for supervised learning tasks such as image classification it has been shown that a good measure of complexity is the intrinsic dimension of the manifold of the images of the dataset \cite{ansuini2019intrinsic}, here the situation is more unclear and a systematic study from DMFT seems possible.
\item \emph{Possible interplay between Hebbian and FORCE training.--} It is well known that standard RNN can learn a task only if the level of chaos is within some working range (which may be dependent on the complexity of the task) \cite{sussillo2009generating,sussillo2009learning}. The same happens also if we use the dynamical system in Eq.~\eqref{dyn_sys}. This is reasonable: if the level of chaos is too small, the endogenous dynamics is not sufficient to sustain the activity needed to produce a target function. Conversely, if the level of chaos is {too} strong, the system experiences wild fluctuations which prevent training. It would be very interesting if one could use Hebbian training as a way to tune the level of chaos during FORCE learning, in such a way that the learning task could be performed optimally. 
\item \emph{Hebbian learning: node perturbation and variants.--} FORCE learning, while being very effective, lacks of biological plausibility. For example, the algorithm relies on the computation of the matrices $P(t)$ which needs to be done off-line. It is clear that if one wants to use RNNs to model biological neural networks, it is crucial to engineer training strategies that are closer to be biologically plausible. In recent years, such line of research has been started and a few training strategies with varying degree of biological plausibility have been proposed, see \cite{fiete2006gradient, fiete2007model, miconi2017biologically}. A number of them is based on the use of an eligibility trace to solve the credit assignment problem. While in some cases there is a clear theoretical foundation for the working mechanisms of the algorithm \cite{fiete2006gradient}, in others, the working principles are less understood and very limited \cite{miconi2017biologically}. A possible perspective is to try to adapt and use these training strategies in the context of the models we have been studying in this work.
\item \emph{The high-dimensional competitive limit of linear readout units.--} We generalized our framework to the case in which there are many linear readout units. They are not directly interacting (there is no synaptic connection between them) but their interaction is mediated by the dynamical system itself. In this setting, there are two interesting perspectives to be investigated. On the one hand, it would be interesting to understand how two readout units can be trained to perform competitive tasks (which are tasks that are mutually exclusive to some degree) and what is the resulting dynamics.
The other interesting limit to look at is when the number of the readout units is sent to infinity (but after the thermodynamic limit of the dynamical system itself). This would be an approximation for the situation in which the size of the central neural network is huge as compared to the peripheric neural network (and it is the same setting that one encounters in low dimensional activities such as motor control).
\item \emph{High-dimensional optimal control and generative modeling.--} In the previous sections, we have refereed to the endogenous drive term in Eq.~\eqref{dyn_sys} as a chaotic noise, see Eq.~\eqref{micro_xi}. An interesting perspective is to use this out-of-equilibrium noise as a bath to drive the readout units to explore target probability distributions. This would be the same strategy as in \cite{song2020score}. Given that the process of biasing a stochastic process to sample a given probability distribution can be recast into an optimal control problem \cite{fleming1977exit}, it is clear that this perspective is directly linked to high-dimensional version of optimal control \cite{urbani2021disordered} and the key point will be to control the statistics of the readout weights. It is also important to note that in this case the goal of the network is not to suppress chaos as in the learning tasks we have discussed in this work, but rather to control it.
\item \emph{Spiking neural networks.--} This work has focused on a random high-dimensional chaotic system as a simplified and abstract model of a RNN. It would be interesting to investigate if this work can be generalized to spiking dynamics to model spiking neural networks \cite{izhikevich2007dynamical}.
\end{enumerate}
Therefore, we believe that this work opens a set of interesting directions that we plan to explore in forthcoming works.

%\bibliography{HS}

\end{document}